\documentclass[%
 reprint,
superscriptaddress,
nofootinbib,
 amsmath,amssymb,
 aps,
]{revtex4-2}

\usepackage{graphicx}
\usepackage{dcolumn}
\usepackage{bm}
\usepackage{hyperref}

\usepackage{titlesec}
\titlespacing*{\section}{20pt}{0.9\baselineskip}{0.55\baselineskip}

\begin{document}
\preprint{APS/123-QED}

\title{Radiative Signatures from Warp Drives Traveling Through the Earth's Atmosphere}

\author{Shaun David Brocus Fell}
 \email{shaun@appliedphysics.org}
\affiliation{%
 Applied Physics, PBC\\
 477 Madison Avenue\\
 New York, NY 10022, USA
}%

\author{Abraham Loeb}
 \email{aloeb@cfa.harvard.edu}
\affiliation{%
 Applied Physics, PBC\\
 477 Madison Avenue\\
 New York, NY 10022, USA
}%
\affiliation{
Astronomy Department, Harvard University\\
60 Garden St., Cambridge, MA 02138, USA
}%

\date{\today}

\begin{abstract}
We investigate the observable signatures of zero-ADM-mass warp-drive spacetimes traversing Earth’s atmosphere. Numerical simulations indicate that an aircraft-scale spacetime bubble moving at relativistic velocities would have a pronounced observational signature, where interaction with the atmosphere can produce luminosities exceeding one terawatt. The signature of a spacetime bubble at rest or moving at low velocity relative to the Earth would not generate such extreme luminosities. These results establish observational constraints on spacetime-based propulsion operating within the terrestrial environment and provide a framework for identifying potential high-velocity signatures. In particular, a warp drive traveling through the atmosphere at speeds exceeding approximately 10\% of the speed of light would produce a unique brilliant glow.
\end{abstract}


\maketitle

\section{Introduction} \label{sec:introduction}

Recent years have seen a substantial uptick in theoretical and numerical research into exotic gravity-mediated propulsion concepts, most notably the warp drive~\cite{lentz2021breaking, bobrick2021introducing, fell2021positive, santiago2022generic, schuster2023adm, helmerich2024analyzing, helmerich2024warpfactory, fuchs2024constant, celmaster2025violations, barzegar2026general}. These studies have ranged from purely theoretical analyses of constructing physically-realizable warp drive geometries to the detectability of such highly non-linear spacetimes in the cosmos~\cite{clough2024warpwaveforms, barcelo2022aerodynamics, bian2022obstacles,lentz2024emissions}. 

Several studies explored detectability of warp drive spacetimes by analyzing either gravitational radiation~\cite{clough2024warpwaveforms, sellers2022searching, kuwahara2023development}, emissions~\cite{lentz2024emissions, clark1999null}, or optical lensing~\cite{anderson2011ray, muller2012detailed}. For example,~\citet{clough2024warpwaveforms} analyzed the gravitational radiation emitted from an Alcubierre-like metric undergoing collapse.~\citet{sellers2022searching} showed that present day gravitational wave detectors are sensitive to rapidly and/or massive accelerating objects, which would include non-zero ADM mass accelerating warp drive spacetimes.~\citet{muller2012detailed} analyzed the geodesics of the Alcubierre spacetime to visualize the lensing of the light emitted from distant sources.~\citet{bian2022obstacles} investigated how warp drives interact with the interstellar medium. 

All of the studies so far have only analyzed how warp drives behave in interstellar space. Either warp drives interact with interstellar material, they emit gravitational radiation, or lens distant light. However, how warp drives behave in the Earth's atmosphere is still an open question. If a warp drive enters at relativistic velocities, it violently interacts with the upper atmosphere resulting in extreme Mach numbers, shocks, pair production, ionization, and extreme luminosities. In the present study, this interaction is investigated numerically.

Reports on Unidentified Anomalous Phenomena (UAP) near Earth ~\cite{Watters,Knuth} triggered speculations that they might be associated with warp drives. The present study is performed with this association in mind and observational signatures are presented within this context, following~\citet{Loeb}.

As a representative embodiment, the metric of a warp drive spacetime is taken to be the Alcubierre metric. There are numerous warp drive metrics in the literature that could be chosen, with varying asymptotic conditions and geometric properties, however the shock structure and luminosities will be roughly the same for the class of zero ADM mass spacetimes, since the Rankine-Hugoniot conditions are largely dependent on the relative velocity between the warp bubble comoving frame and the fluid frame and not the fine structure of the geometry. A caveat to this is that the Alcubierre metric has very different asymptotic structure than conventional spacetimes. Outside a given spatial scale in the Alcubierre spacetime, the Riemann curvature is exponentially suppressed, in contrast to more conventional asymptotically flat spacetimes, such as Schwarzschild. This implies the interaction has the potential to behave qualitatively the same as a solid ballistic object incident on the atmosphere. A spacetime with different asymptotic properties is likely to behave qualitatively differently.

The relevant equations for this interaction are the covariantly conserved flux equations written in the Valencia formulation~\cite{Servignat_2023} with an appropriate shock capturing scheme layered on top. Moreover, since the characteristic energy scale is above the electron dissociation scale, electrons will be stripped from their atoms, forming a non-equilibrated electron-ion gas. This motivates the inclusion of a two-temperature model of the gas, where the electrons and ions carry their own thermal distributions and a potential thermal coupling between them. 

Due to the high relative velocity of the fluid and the warp bubble, there is a large characteristic energy scale of the interaction, which can be estimated using basic properties. In front of the bubble, an on-axis stagnation point is formed where the fluid comes nearly to a halt, thus all the relative kinetic energy is transformed into thermal excitation energy. The maximum energy dumped into the fluid occurs within this region and can be approximated by equating the relative kinetic energy with the thermal energy. For a typical bubble velocity of $v_s = 0.5c$, this results in the typical thermal energy on the scale of $144\, \mathrm{MeV}/\text{nucleon}$, well above the electron pair production point and slightly past the pion threshold~\cite{ParticleDataGroup}. The additional particle production will change the energy transfer rate of the gas. How fast that energy transfer happens depends on the production rate, which in turn depends on the overall density of the gas.

It will turn out these physical quantities don't meaningfully affect the dynamics, so that the simulation can be run without advanced radiative transfer additions or additional degrees of freedom in the equation of state. The shocked gas is optically thin across all simulated parameters, so photons efficiently escape the shock zone. However, the hydrodynamical timescale is much shorter than the radiative cooling time, rendering radiative losses negligible. 

The high Mach number limit also corresponds, to within an $O(1)$ constant, to the cold-gas limit, $p/\rho \ll 1$. Therefore, for higher Mach numbers, the simulation better predicts the cold-temperature limit, even if the simulation is performed in a warm regime.

The rest of the paper is structured as follows: Section~\ref{sec:theory} lays out the theoretical foundation for the study, including the relevant dynamical equations and equation of state, basic fluid microphysics, expected fluid behavior, and chosen simulation code. Section~\ref{sec:results} discusses the simulation itself and the results of the simulation runs. Section~\ref{sec:discussion} discusses the simulation results. Section~\ref{sec:conclusion} wraps up the paper with an outlook, suggested future studies, and impact on current political topics.

\section{Theory} \label{sec:theory}

\subsection{Metric}
The metric of the spacetime is taken to be the Alcubierre spacetime~\cite{alcubierre1994warp}, where the line element takes the form
\begin{align} \label{eq:alcubierre}
    ds^2 =& -dt^2 + (dx - v_s f(r_s) dt)^2 + dy^2 + dz^2\, ,\\ \nonumber
    r_s =& \sqrt{(x - x_s)^2 + y^2 + z^2}\, ,
\end{align}
where $f$ is the shape function
\begin{equation}
    f(x) = \frac{ \tanh{\left(\sigma (x + R)\right)} - \tanh{\left(\sigma ( x - R)\right)}}{2 \tanh{\left(\sigma R\right)}}\, ,
\end{equation}
where $\sigma$ is the inverse thickness of the bubble wall and $R$ is the radius of the bubble. The entire fluid-geometry coupling is thus wall-localized, since the only nonzero metric derivatives occur at radius $R$ in a region of radial thickness $1 / \sigma$.

The metric is assumed static over the course of the simulation. Moreover, the spacetime source term is ignored. This implies that components of the atmospheric fluid can pass cleanly through the highly nonlinear volumes of the region, which would normally not occur in a physical setup. However, this exclusion of the Alcubierre source has little impact on the shock properties. For example, the stagnation zone sits outside the bubble wall, see Sec.~\ref{subsec:stagnationandjumpconditions}. The shock front sits even further away than the stagnation zone. This greatly simplifies the simulation code without a meaningful impact on the results, since the matter sector consists solely of the interacting fluid without any backreaction.

\subsection{Dynamical Equations} \label{subsec:dynamicalequatios}
The choice of the form of the dynamical equations dramatically impacts the convergence properties of the solution. The Lax-Wendroff theorem gives insight into what choice should be made, which states \emph{if a conservative numerical scheme for a hyperbolic system of conservation laws converges, then it converges towards a weak solution}~\cite{laxwendroff}. Thus, the choice of dynamical equations should be written in a conservative form, where the dynamical variables are conserved with respect to given flux laws. 

Moreover, the defining feature of the interaction is the formation of a high-velocity shock front, forming a discontinuity in the conserved variables. Therefore, the correct choice is a conserved shock-capturing scheme. For the current study, a high-resolution shock-capturing Godunov method is employed, which constructs the primitive variables from the conserved variables, solves an approximate Riemann problem on the cell faces, and updates the state variables conservatively~\cite{Mart1999}. Instead of re-implementing all of this, Athena++ is chosen for its native implementation of general relativistic hydrodynamics~\cite{Stone2020}. 

The primitive variables are the standard fluid quantities: the density $\rho$, specific internal energy $\epsilon$, and pressure $p$. These come directly from the perfect fluid form of the stress-energy tensor describing the interacting fluid\footnote{This perfect fluid describes the material the non-linear spacetime is interacting with and not the stress-energy tensor generating the spacetime curvature.},
\begin{align}
    \mathcal{T}^{\mu \nu} &= \rho h u^{\mu} u^{\nu} + p g^{\mu \nu} \\ \nonumber
    h &= 1 + \epsilon + \frac{p}{\rho}\; ,
\end{align}
where $h$ is the specific enthalpy.

The conserved variables are then precisely the projections of the covariant currents onto the normal observer $n_{\mu} = (-1,0,0,0)$ and onto the hypersurfaces. These form the tuple
\begin{align}
    D &= \rho W \, ,\\
    S_j &= \rho h W^2 \tilde{v}_j \, ,\\
    \tau &= \rho h W^2 - p - D\, ,
\end{align}
where $W = - n_{\mu} u^{\mu}$, with hypersurface normal $n^{\mu}$ and fluid four-velocity $u^{\mu}$, and the Eulerian-frame velocity is $\tilde{v}$, which is related to the fluid velocity $v^i = \frac{u^i}{u^0}$ via $v^i = \tilde{v}^i - \beta^i$. Here, $\beta$ is the shift vector for the Alcubierre metric Eq.\eqref{eq:alcubierre} and is equal to the warp bubble velocity in the direction of motion.

In these variables, the first-order flux-conservative equations take the form~\cite{Font2008}
\begin{equation} \label{eq:dynamical_equations}
    \partial_t\left(\sqrt{\gamma} \vec{U}\right) + \partial_i \left( \sqrt{-g} \vec{F}^i \right) = \sqrt{-g} \vec{S}\, ,
\end{equation}
where $\vec{U} = (D, S_j, \tau)$, the flux is $\vec{F}^i = (D v^i, S_j v^i + p \delta^i_j, \tau v^i + p \tilde{v}^i)$ and is advected by the coordinate velocity $v^i$, and the purely-geometric source term is $\vec{S} = (0, S_i \partial_j \beta^i, S_{ij} K^{ij})$, with $K_{ij} = -\partial_{(i} \beta_{j)}$. Note that the conserved $D$ has zero source, realizing the baryon mass conservation constraint. 

\subsection{Equation of State}
To close Eqs.~\eqref{eq:dynamical_equations}, an equation of state (EoS) needs to be chosen. The EoS sets the sound speed $c_s$ which in turn sets the Riemann wavespeed and thus the timestep of the simulation. The EoS therefore doesn't just set the physics, it also determines the stability of the simulation.

The EoS needs to be chosen that is consistent with the target system and the physics. An ideal gas law with adiabatic index $\Gamma$ would be a simple intuitive choice, but it is thermodynamically inconsistent. The gas needs to be able to handle cold and relativistically hot states. A fixed adiabatic index cannot supply that. The relativistic constraint that needs to be satisfied by the EoS is the Taub inequality~\cite{taub1948}
\begin{equation}
    (h - \Theta)(h - 4\Theta) \geq 1 \quad \text{where} \quad \Theta \equiv \frac{p}{\rho}\, .
\end{equation}

Instead, a Taub-Mathews (TM) EoS is taken, as it interpolates $\Gamma$ from the non-relativistic limit of $\Gamma = 5/3$ to the relativistic limit $\Gamma = 4/3$ and approximates very well a single-species Synge gas without requiring expensive Bessel function evaluations~\cite{MignoneMcKinneyJonathan, Mathews1971}. The TM EoS takes the form
\begin{align}
    h(\Theta) &= \frac52 \Theta + \sqrt{\frac94 \Theta^2 + 1},\\ \nonumber 
    c_s^2 &= \frac{ \Theta}{3h} \frac{5h - 8 \Theta}{h - \Theta}\,.
\end{align}
In the cold limit, $\Theta\rightarrow0$, $h \rightarrow 1 + \frac52 \Theta$ and $c_s^2 \rightarrow \frac53 \Theta$, while in the relativistic limit, $\Theta \rightarrow \infty$, $h \rightarrow 4 \Theta$ and $c_s^2 \rightarrow \frac13$.

\subsection{Recovering the Primitives} \label{subsec:primitives}

The governing dynamical equations evolve directly the conserved quantities $\vec{U} = (D, S_j, \tau)$, but the fluxes, sources, and EoS all require the primitive variables $\vec{P} = (\rho, v^i, p)$. Every step and every discrete cell of the simulation must therefore compute the mapping $\vec{U} \rightarrow \vec{P}$, henceforth called the conservative-to-primitive (con2prim) problem. For a general EoS, there is no closed form so implicit root-finding is required. Moreover, the solution to the con2prim problem dictates whether the simulation is even possible. 

The native scheme to Athena++ is the Newman-Hamlin scheme~\cite{Newman2014}, which solves a 1-D fixed-point iteration on $p$. The pressure update is
\begin{equation}
    p_{\text{new}} = \frac{1}{8} \left( 5 \omega_{\text{gas}} - \sqrt{9 \omega_{\text{gas}}^2 + 16 \rho^2} \right)\, ,
\end{equation}
where $\omega_{\text{gas}} = \frac{a}{W^2}$ and $a \equiv \tau + D + p$. This is iterated until convergence on $p$. This scheme is derivative-free. However, in the current study, it fails as a naive inversion scheme. The first failure is that the update subtracts near-equal $O(\epsilon)$ quantities, so successive iterations jitter by an absolute $\sim$ few $\epsilon_{\rm mach} \epsilon$, where $\epsilon_{\rm mach}$ is machine epsilon, independent of how small $p$ is. The stock convergence test is a relative-only check $|\delta p| < \text{tol}\times p$, with tol$=10^{-12}$. This convergence fails whenever $\Theta \leq 10^{-4}$. Every such cell then runs to $\texttt{max\_iterations}$ while physically converged, and the stock accept-best-iteration path chooses whichever garbage it stopped on. Instead, an absolute term is included in the check
\begin{equation}
    | \Delta p| < \text{tol} \times p + 20 \epsilon_{\text{mach}} \epsilon\, ,
\end{equation}
with $\texttt{max\_iterations}$ taken to be $30$ and a relative per-iteration floor to be $p_{\text{min}} = 10^{-14} \epsilon$. If a cell still fails to converge, an explicit status is emitted. 

The failing cells are, however, precisely the cells that never need $\tau$. No dissipation acts on the cells because they are adiabatic, so each fluid element conserves its specific entropy along its own worldline, $u^{\mu}\partial_{\mu}S = 0$, where $S$ is the entropy per baryon. Their thermal state has not changed since they entered the domain, so it can be carried instead of recovered. For the TM EoS, the adiabat is
\begin{equation}
    \sigma_{TM} = \frac{p}{\rho^{5/3}} (h - \Theta)\, ,
\end{equation}
which converges to the ideal gas adiabat $K = \frac{p}{\rho^{5/3}}$ as $\Theta \rightarrow 0$. $\sigma_{TM}$ is a per-baryon label, so it travels with the conserved rest-mass density $D$. Defining the entropy density $s = D \sigma_{TM}$, the conservation equation yields
\begin{equation}
    \partial_t s + \partial_i (s v^i) = 0\,.
\end{equation}
Then updating the pressure from a carried $\sigma_{TM}$ involves a Newton iteration starting from the cold-limit seed $p_0 = \sigma_{TM} \rho^{5/3}$. The simulations show that three steps is sufficient to drive the residual below $10^{-13}$ throughout cold, kinetic-dominated cells. 

The above scheme is the fallback branch of the con2prim inversion algorithm. The deciding layer that determines which scheme to use is the ENZO dual-energy formalism~\cite{Bryan_2014}. Instead of evolving the second internal-energy field used in ENZO, the auxiliary variable is the advected entropy invariant above. 

Thus the con2prim solution logic is as follows:
\begin{itemize}
    \item Read the carried adiabat $\sigma_{TM}$ from the conserved scalar.
    \item Run the Newman-Hamlin iteration.
    \item If the iteration exits with an unhealthy status, go straight to the entropy advection layer.
    \item Otherwise, compute the selector scalar $\chi = {p_{eng}}/{\rho(W-1) + p_{eng}}$. $p_{eng}$ is the energy-recovered pressure.
    \item If the selector is above a given threshold, trust $p_{eng}$ and return.
    \item If the selector is below a given threshold, the cell is deemed cold kinetic-dominated and the entropy advection layer is applied.
\end{itemize}
In the simulations, the threshold is taken to be $10^{-3}$. A caveat to this scheme is that the entropy advection layer is physically false to use across the shock. Nonetheless, the energy-based inversion correctly runs in this region anyways.

\subsection{Stagnation and Jump Conditions} \label{subsec:stagnationandjumpconditions}

The post-shock energy per nucleon is fixed by the relativistic Rankine-Hugoniot (RH) jump conditions and $v_s$ alone. Steady-front flux continuity of baryon number, momentum, and energy yield the jump conditions
\begin{equation}
    [ \rho W v_s] = [\rho h W^2 v_s^2 + p] = [\rho h W^2 v_s] = 0\, ,
\end{equation}
where the square brackets denote the difference across the shock front, $[X] \equiv X_{2} - X_{1}$, where $X_2$ is the post-shock quantity and $X_1$ is the upstream, pre-shock quantity. On the nose of the warp bubble, the post-shock fluid comes almost to a complete halt, forming a stagnation zone. Within this zone, all the kinetic energy of the fluid is deposited into thermal energy, thus yielding a characteristic energy scale of the shock itself. By simply equating the thermal energy to the relativistic kinetic energy, one finds the temperature scale as
\begin{equation}
    k_{B} T_{\text{stag}} \sim (W_1 - 1) m_u c^2\, ,
\end{equation}
where $W_1$ is the Lorentz factor of the upstream fluid and $m_u$ is the atomic mass. For example, at a relative velocity of $v = 0.5c$, the stagnation point has characteristic thermal energy on the scale of $k_B T_{\text{stag}} \approx 144\, \mathrm{MeV}$, which is well above the electron pair production threshold of $\sim 1.022\, \mathrm{MeV}$. Moreover, this is slightly above the threshold for neutral pion production, which is around $135\, \mathrm{MeV}$. At $v_s = 0.75$, the characteristic thermal energy is well-beyond the pion thresholds.

The compression ratio $\frac{\rho_2}{\rho_1}$ follows directly from the RH conditions, yielding
\begin{equation}
    \frac{\rho_2}{\rho_1} = W_1 \frac{v_{\mathrm{sh}} - v_1}{v_{\mathrm{sh}}}\, ,
\end{equation}
where $v_{\mathrm{sh}}$ is the shock front speed and $v_1$ is the incoming upstream fluid speed. These quantities are relative to the bubble comoving frame.

Another important quantity is the standoff distance from the bubble exterior 'wall' to the shock cone itself. The bubble here is taken to be a blunt solid object of radius $R$ and a shock standoff distance of $R_s = R + \Delta$. Where the detached shock sits follows directly from mass conservation in the compressed layer. In the steady state, the upstream mass the cap captures cannot accumulate. Instead, it drains out tangentially through the annular cross-section of a thin layer. The thin layer is taken to have density $\rho_2$ (post-shock density) and tangential drain speed $u_t \sim C v_s \theta$, where $\theta$ is the half-angle of the cap itself. The constant $C \sim O(1)$ is the stagnation point velocity-gradient coefficient. The conservation equation and cross-section are then
\begin{align}
    \rho_1 v_s A_{\text{cap}} &= \rho_2 u_t A_{\text{layer}} \\
    (A_{\text{cap}}, A_{\text{layer}}) &= (\pi R^2 \theta^2, 2 \pi R \theta \Delta)\, ,
\end{align}
so that
\begin{equation} \label{eq:shockstandoff}
    \frac{\Delta}{R} = \frac{\rho_1}{2 C \rho_2}\, ,
\end{equation}
which is inversely related to the compression ratio. A larger compression ratio corresponds to a tighter layer, as expected. 

The RH jump conditions can also be used to show explicitly how high Mach numbers correspond to independence of the shock properties with the upstream temperature $\Theta_1$.  In the comoving frame, the three RH equations become
\begin{align}
    \text{Energy} &: v_{\mathrm{sh}}(\rho_2 h_2 - p_2 - \rho_1 h_1 W_1^2 + p_1) = - \rho_1 h_1 W_1^2 v_1\, , \\
    \text{Momentum} &: \rho_1 h_1 W_1^2 v_1(v_{\mathrm{sh}} - v_1) = p_1 - p_2 \, ,\\ \label{eq:compratio}
    \text{Mass} &: \frac{\rho_2}{\rho_1} = W_1 \frac{v_{\mathrm{sh}} - v_1}{v_{\mathrm{sh}}}\,.
\end{align}

The post-shock temperature can then be written exactly as
\begin{equation}
    \Theta_2 := \frac{p_2}{\rho_2}= \frac{v_{\mathrm{sh}} \Theta_1}{W_1(v_{\mathrm{sh}} - v_1)} - h_1 W_1 v_1 v_{\mathrm{sh}}\,.
\end{equation}
In the cold-limit as $\Theta_1 \rightarrow 0$, this reduces to
\begin{equation}
    \Theta_2 \sim - h_1 W_1 v_1 v_{\mathrm{sh}}\, ,
\end{equation}
where the convention is that the incoming upstream fluid has negative velocity $v_1$, thus making $\Theta_2$ positive. Substituting the mass and momentum relations into the energy relation, one finds the enthalpy post-shock as
\begin{equation}
    h_2 = h_1 W_1 ( 1 - v_1 v_{\mathrm{sh}})\, ,
\end{equation}
and using the internal energy relation $\epsilon = h - 1 - \Theta$, the post-shock downstream internal specific energy is
\begin{align}
    \epsilon_2 &= h_1 W_1 - 1 - \frac{p_1}{\rho_2}\\
    &= h_1 W_1 - 1 - \frac{p_1 v_{\mathrm{sh}}}{\rho_1 W_1 (v_{\mathrm{sh}} - v_1)}\, ,
\end{align}
which depends entirely on the upstream temperature $\Theta_1$ only through the EoS, and only weakly. In the cold-limit, one finds the expansion
\begin{equation}
    \epsilon_2 = (W_1 - 1) \left( 1 + a_1 \Theta_1 + O(\Theta_1^2)\right)\, ,
\end{equation}
where the linear coefficient is
\begin{equation}
    a_1 = \frac{15W_1^4 v_1^2 + 5W_1^4 - 7W_1^2 + 2}{2W_1(W_1-1)\left(3W_1^2 v_1^2 + W_1^2 - 1\right)}\,.
\end{equation}
Since $a_1>0$, the warm limit approaches the cold limit from above. To see how this scales with Mach number, consider the deviation
\begin{equation}
    \delta = \frac{\epsilon_2}{W_1-1} - 1 = a_1 \Theta_1 + O(\Theta_1^2)\,.
\end{equation}
Then rewrite $\delta = C_A \frac{1}{\mathcal{M}^2}$, where $C_A =\frac35 a_1 v_1^2$. By simply computing the two limits $v_1 \rightarrow 0$ and $v_1\rightarrow 1$, one finds that $C_A$ is bounded from below and above by an $O(1)$ constant. Precisely, $C_A|_{v_1\rightarrow 0} = 2.7$ and $C_A|_{v_1\rightarrow1} = \frac32$. Thus, $\delta$ behaves like $\sim \frac{1}{\mathcal{M}^2}$ with an order 1 coefficient. This cold-limit/high Mach-number equivalence also holds for the other fluid quantities. This implies that large Mach numbers correspond to the cold upstream limit to within an order 1 coefficient. This implies simulations run at high Mach numbers become agnostic to the upstream temperature $\Theta_1$. Hence, the simulation results from warm runs $\Theta_1 \lesssim 1$ apply also to colder runs, for example $\Theta_1 \sim 10^{-13}$, when $\mathcal{M} \gg 1$. This enables one to produce results that apply to the Earth's own atmosphere without needing expensive arbitrary precision. One need only to execute simulations for much more achievable temperatures such as $\Theta_1 \sim 10^{-3}$ and take the large Mach limit, which is easily achievable with warp drive spacetimes.

\subsection{Adaptive Mesh Refinement}
The physics is largely confined to two thin, \emph{a priori} unknown-in-positions sheets. The first is the metric source region and the second is the standing bow shock it drives. A block AMR condition is enforced that pins the warp bubble wall to the finest level and tracks the shock adaptively with a flag that fires on the shock feature. The flag itself is the L\"ohner normalized second difference~\cite{Lhner1987}. 

The fluid-geometry coupling is dependent on the derivative of the Alcubierre shape function, $f^{\prime}(r)$, which peaks at $r=R$ with width $\frac{1}{\sigma}$. The AMR refinement condition refines every block that overlaps with the shell, ensuring the non-linear aspects of the geometry are fully resolved with the highest resolution.

A refinement flag for the shock requires more care, due to its dynamical nature. The refinement flag must be dimensionless, scale-invariant, and noise-immune. In this study, the indicator for the flag is the curvature of the fluid Eulerian energy $\partial^2 E$, made dimensionless by the variation $|\partial E|$. For a given cell $ijk$, $E_0 = E_{ijk}$ and $E_{\pm} = E$ at its two nearest cell-centered neighbours along the direction d, where $i \pm 1$ for d=x, $j \pm 1$ for $d = y$ and $k \pm 1$ for d = z. This defines a stencil $\{E_-, E_0, E_+\}$ for each grid direction. The undivided second difference for the numerator is then
\begin{equation}
    \Delta_d^2 E = E_+ - 2E_0 + E_- = \delta^2 \partial^2 E + O(\delta^4)
\end{equation}
for cell spacing $\delta$.  The ratio $\frac{\delta}{2} \frac{ |\partial^2 E|}{|\partial E|}$ is dimensionless and scale invariant. Then the L\"ohner noise filter $\epsilon_f (|E_+| + 2 |E_0| + |E_-|)$ is added to the denominator to make the indicator flag noise-resistant. Across all directions, the total indicator flag is then
\begin{equation}
  \chi_L = \left[ \frac{\sum_{d} (\Delta^d)^2}{\sum_d (\eta^d)^2} \right]^{1/2}\, ,
\end{equation}
where
\begin{align}
  \Delta^d \equiv& E^d_+ - 2E_0 + E^d_- , \\
  \eta^d \equiv& |E^d_+ - E_0| + |E_0 - E^d_-|\\
           & + \epsilon_f \left( |E^d_+| + 2|E_0| + |E^d_-| \right)\,.
\end{align}
$\chi_L$ is therefore bounded by $[0,1]$. The refinement condition for the AMR grid is then
\begin{equation}
    \texttt{return} = \begin{cases}
        \texttt{refine} & \text{on wall or }  \chi_L > \chi_{\text{ref}} \\
        \texttt{reduce} & \text{not on wall and } \chi_L < \chi_{\text{der}}\\
        0 & \text{otherwise}
    \end{cases}
\end{equation}
A given block takes the maximum of $\chi_L$ across all its cells, which gets fed into the refinement condition above. The implemented reference is $\chi_{\text{ref}} = 0.8$ and $\chi_{\text{der}} = 0.3$ and the noise amplitude is $\epsilon_f = 0.01$.

\subsection{Derived Thermodynamics}

\subsubsection{Optical Depth} \label{subsubsec:opticaldepth}
A large amount of the science does not need explicit evolution during the simulation. To high accuracy, most of the thermodynamic quantities are just algebra or calculus on the primitives through the EoS. One of the most important ones is the optical depth of the fluid. For the discretized grid, the Thomson optical depth is
\begin{equation}
    \tau_T = \sigma_T \sum_{x_i \geq R} \left(n_e(j_0, i) +n_+(j_0, i) +n_-(j_0, i) \right) dl\, ,
\end{equation}
where $n_+$ and $ n_-$ are the pair-produced particle densities, $n_e = \frac{\rho_{phys}}{m_u} \frac{Z}{A}$ is the electron density,  $m_u$ is the atomic mass, $\frac{Z}{A} = 0.5$ is the ratio of the atomic number to mass number of the fluid constituents, here taken to be predominantly oxygen and nitrogen, and $\sigma_T$ is the Thomson scattering cross-section. The pair-produced particle densities are estimated from the photon emission. The computation runs on the final output of the simulation. The $2$-tuple $(j,i)$ indexes the $(y,x)$ cells with cell-centers $x_i$. The optical depth is computated on the stagnation line, which happens to coincide with the $x$-axis and $j_0$ is the cell row closest to this axis. 

The sum begins at the bubble wall radius $R$ and runs to the domain boundary at $\mathfrak{U}R$ (See Sec.~\ref{subsubsec:thecode}), so the ray crosses the bow shock and continues through the undisturbed upstream fluid. Both regions therefore contribute. For example, at $\Theta_1 = 10^{-3}$, $v_s = 0.5c$, and $R = 100$ m, the shocked layer between $x = 100$ m and $x = 183$ m contributes $3.4 \times 10^{-6}$, while the remaining upstream column contributes $4.7\times 10^{-6}$, for a total of $8.1 \times 10^{-6}$. A caveat to the upstream contribution is that the expression for $n_e$ assumes the fluid is fully ionized. This is well satisfied in the shocked layer, where the thermal energy sits far above every ionization energy, but the cold ambient is largely neutral and its true electron column is negligible. The upstream contribution is thus an overestimate. Across $v_s = 0.1c$ to $0.75c$, $\tau_T$ stays between $7.8\times10^{-6}$ and $8.2\times10^{-6}$, while the compression ratio over the same range varies from $2.97$ to $5.40$. The shocked-layer value is therefore taken as the physical optical depth, and the full-ray value as an upper bound.

The pair densities are estimated from the local photon emission by balancing pair creation against annihilation, with the buildup capped by the time a single fluid element stays in the shell. The non-equilibrium treatment is thus necessary here. For the same fiducial parameters above, the balance gives $n_{\pm} \approx 2\times10^{9}\ \mathrm{m}^{-3}$ against an electron density of $n_e \approx 7.6\times10^{20}\ \mathrm{m}^{-3}$, so the pair contribution to the optical depth is $\tau_{\text{pair}} \approx 3\times10^{-17}$, roughly eleven orders of magnitude below the baryonic column. Note that the photon field used to obtain this is the freely escaping one, which presupposes the flow is optically thin. The resulting $\tau_{\text{pair}} \ll 1$ recovers that assumption.

In the relativistic context, the appropriate scattering cross-section is the Klein-Nishina cross-section. However, the relativistic cross-section is less than or equal to the non-relativistic approximation, so for the purposes of showing optically thin flow, the non-relativistic form is sufficient. Across all the simulation parameters, the optical depth  stays below $10^{-5}$, well into the optically thin range. 

\subsubsection{Interaction Timescale} \label{subsubsec:interactiontimescale}
The second defining quantity that determines whether the derived quantities can be computed in post-processing is the relation between the interaction timescale of the fluid constituents relative to the flow timescale. The ratio of the interaction timescale to the flow timescale is
\begin{equation}
    \frac{t_{ei}}{t_{flow}} = \sqrt{ \frac\pi2 } \frac{m_i}{m_e} \frac{ (\Theta_e + \Theta_i)^{3/2} v_s}{Z \ln \Lambda (n_e \sigma_T R)}\, ,
\end{equation}
where $m_i$ is the ion mass, $m_e$ is the electron mass, $\Theta_e$ is the electron temperature in units of the electron rest energy, $\Theta_i$ is the ion temperature in units of the ion rest energy, and $\ln \Lambda$ is the Coulomb logarithm. For typical parameters in the simulation, $\frac{t_{ei}}{t_{flow}} \sim 10^7$, thus the interaction timescale of the electrons and ions vastly exceeds the typical flow timescale. Therefore, the two species don't equilibrate and thus stay decoupled across the entire shell, implying the gas carries two internal energies,
\begin{align}
    \partial_t \tau_i + \partial_i F^i_{\tau_i} &= (1 - \xi)Q_{sh}\\
    \partial_t \tau_e + \partial_i F^i_{\tau_e} &= \xi Q_{sh}\, ,
\end{align}
where $Q_{sh}$ is the rate at which bulk kinetic energy is irreversibly converted to internal energy,  and $\xi$ is the fraction delivered to electrons, which is set by the kinetic microphysics~\cite{Shapiro1976}. Radiative losses are omitted, since the cooling time far exceeds the flow time.

Instead of computing $T_e$ directly, which loses precision in the cold-upstream cancellations, a gas-entropy tracer is advected as a pure scalar in the same form as Sec.~\ref{subsec:primitives}. The accumulated dissipation $\int Q_{sh} dt$ is computed from the excess of the actual entropy over the advected tracer. The electron fraction needs to be provided by the microphysics of the system, which is beyond the scope of this study, since the fluid code cannot derive the electron/ion heating fraction at a collisionless shock, which is a kinetic result. The front itself is mildly relativistic ($W_1 \in [1.005, 1.511]$), dominated by electron-ion interactions, and effectively unmagnetized, thus the shock is Weibel-dominated and constrained by kinetic simulations~\cite{Weibel1959, crumley2019kinetic}. Particle-In-Cell simulations of Weibel-mediated electron-ion shocks find the downstream temperature ratio $r = T_e/T_i$ to lie between $1/6$ and $1/2$, which converts to an energy fraction by
\begin{equation}
    \xi \simeq \frac{r}{1+r}, .
\end{equation}
The lower end comes from a mildly relativistic shock at the velocity of interest here, $v \approx 0.75c$, but with a finite magnetization of $\sigma_0 = 0.007$ and $M_A \approx 15$~\cite{crumley2019kinetic}. The upper end comes from unmagnetized shocks but which run at much larger Lorentz factors of $\gamma_{\infty} = 10 - 100$~\cite{spitkovsky2008structure, vanthieghem2022origin}. The shock zone here is unmagnetized like the latter, but only mildly relativistic like the former. Thus $\xi$ is taken to be a bracket, $\xi \in [0.1, 0.5]$, rather than a prediction. For the current study, the typical value of $\xi$ is roughly $0.2$, which is fixed in the post-processing results for the luminosity calculations.

\subsubsection{Pair Production} \label{subsubsec:pairproduction}

The post-shock energy budget exceeds the pair threshold by two orders of magnitude, so electron-positron pairs inevitably form. Whether the pairs stay as tracer objects or cascade into an optically-thick fireball is the single qualitative question about the interaction's appearance.

Thermal pair production switches on when the Wien tail carries photons above the invariant-mass threshold of $2 m_e c^2 \approx 1.022\, \mathrm{MeV}$, or when the characteristic thermal temperature exceeds $k_B T_e \geq 0.511\,  \mathrm{MeV}$. For the post-shock fluid, $k_B T_e = \xi \Theta_2 m_u c^2 \approx 19.2\,  \mathrm{MeV}$ at $v = 0.5c$ and $k_B T_e \approx 0.625\,  \mathrm{MeV}$ at $v = 0.1$.

\subsubsection{Luminosity} \label{subsubsec:luminosity}

Free-free emission from the hot electrons has a frequency-dependent emissivity. For a non-relativistic Maxwellian, the power per unit volume is
\begin{equation}
    \frac{ dE}{dt dV} = \left( \frac{ 2 \pi k_B T_e}{3 m_e}\right)^{1/2} \frac{2^5 \pi e^6}{3 h_p m_e c^3} \bar{g}_B \sum_s Z_s^2 n_s n_e
\end{equation}
where $\bar{g}_B$ is the frequency average of the velocity averaged Gaunt factor, $m_e$ is the electron mass, $h_p$ is the planck constant, and where $E$ is the total energy~\cite{Rybicki1985}. The relativistic enhancement multiplies the emissivity by $(1 + 4.4\times 10^{-10}T_e)$ increase, with $T_e$ in Kelvin, yielding
\begin{align} \label{eq:emissionpower}
    \frac{ dE}{dt dV} = &\left( \frac{ 2 \pi k T_e}{3 m_e}\right)^{1/2} \frac{2^5 \pi e^6}{3 h_p m_e c^3} \bar{g}_B\\ \nonumber 
    & \times\left( 1 + 4.4 \times 10^{-10} T_e\right) \sum_s Z_s^2 n_s n_e 
\end{align}
depending on the temperature of the electron gas, $T_e = \xi \Theta_2 \frac{m_u c^2}{k_B}$. The relativistic enhancement is significant for the high velocities in the current study. The total luminosity is then simply
\begin{equation}
    L = \int \frac{d E}{dt dV} dV
\end{equation}

Typical values range from $10^{13}$ up to $\sim 10^{19}$ watts. The efficiency of the conversion from kinetic energy to total radiated power can be computed straightforwardly. The kinetic power incident on the bubble is the mass flux times the specific kinetic energy. In the bubble's comoving frame, this is
\begin{equation}
    P_{\text{kin}} = (W_1 - 1) W_1 \rho_{phys,\infty} (v_1 c)c^2 \pi R^2
\end{equation}
Typical values for the simulated parameters range from $P_{\text{kin}} \sim 10^{19}$ to $\sim 10^{22}$ watts. This implies the efficiency of the conversion ranges from $10^{-7}$ to $10^{-4}$. This separately confirms the slow radiative cooling compared to the flow timescale. 

The luminosity calculation integrates only over the shock zone, using a temperature cutoff, and subtracts residual background free-free emissions, whether physical or not. For the warm-proxy $\Theta_1 = 10^{-3}$, the ambient atmosphere would in principle also emit its own free-free photons. However, the warm temperatures are only used to determine the much colder regime dynamics $\Theta_1 \sim 10^{-13}$ and thus the ambient emission is unphysical in this regime. Hence, in the luminosity computation, the background is subtracted to estimate the luminosity purely from the hot post-shock state.

\subsubsection{The Code} \label{subsubsec:thecode}

The deployed code heavily utilizes Athena++, an adaptive mesh refinement-based code for magnetohydrodynamical simulations in special and general relativity~\cite{Stone2020}. It fits well for this problem, since the background spacetime is non-linear, stationary, and the only dynamical object is the relativistic unmagnetized fluid interacting with the curved background. The problem space is two-dimensional, exploiting the axial symmetry of the metric and the matter sector.

The simulation grid scales with the size of the warp bubble. The $(x,y)$-directions range over 
\begin{align}
    \text{Dom}(x) &= \left[ - \mathfrak{D}\times R, \mathfrak{U}\times R\right]\\
    \text{Dom}(y) &= \left[ - \mathfrak{T}\times R, \mathfrak{T}\times R\right]
\end{align}
where $\mathfrak{D} = 30$, $\mathfrak{U} = 6$, and $\mathfrak{T} = 10$. The $x$-direction at the coarsest level has $900$ cells, while the $y$-direction has $690$. With $4$ refinement levels chosen for the production runs, this corresponds to the finest refinement level having a resolution of $\Delta x = 0.0025R$ and $\Delta y = 0.0018R$, ensuring that the entire non-linear region is sufficiently resolved in the simulation. The ambient energy density $\rho$ is set to an initial value of $\rho_{\infty} = 1.0$, hence the temperature $\Theta$ sets the pressure term relative to the energy density.

The simulations were executed on Harvard's FASRC computing cluster, using two of their $\texttt{sapphire}$ nodes for each simulation, which house two Intel Xeon Sapphire Rapids CPUs, $990$ GB of RAM, and NDR InfiniBand fabric interconnects.

Athena++ was compiled with OpenMP and MPI parallelization and native SIMD vector intrinsics using the $\texttt{--cflag='-march=native'}$ flag.

\section{Results} \label{sec:results}

For the parameters of the simulation, the upstream temperature (hence the pressure relative to the energy density) takes the values $\Theta_{1} \in \{10^{-3}, 10^{-4}, 10^{-5}, 10^{-6}\}$. These four values will showcase the convergence of the thermodynamical quantities as the upstream temperature becomes extremely small and thus the Mach number increases. The velocity values of the warp bubble are chosen to be $v_s \in \{0.1, 0.25, 0.5, 0.75\}$, covering the weak pair production to the extremely strong $e^-  e^+$ pair production regime. Finally, the warp bubble radius takes values in $R \in \{25, 50, 100\}$ in code units. The code units are taken to be meters and the time units specified such that $c=1$. The warp bubble wall thickness tracks the bubble radius via $\sigma = \frac{5.0}{R}$, which provides a sufficiently flat interior region and well-defined wall boundaries.

\begin{figure*}[tp]
    \centering
    \includegraphics[width=\textwidth]{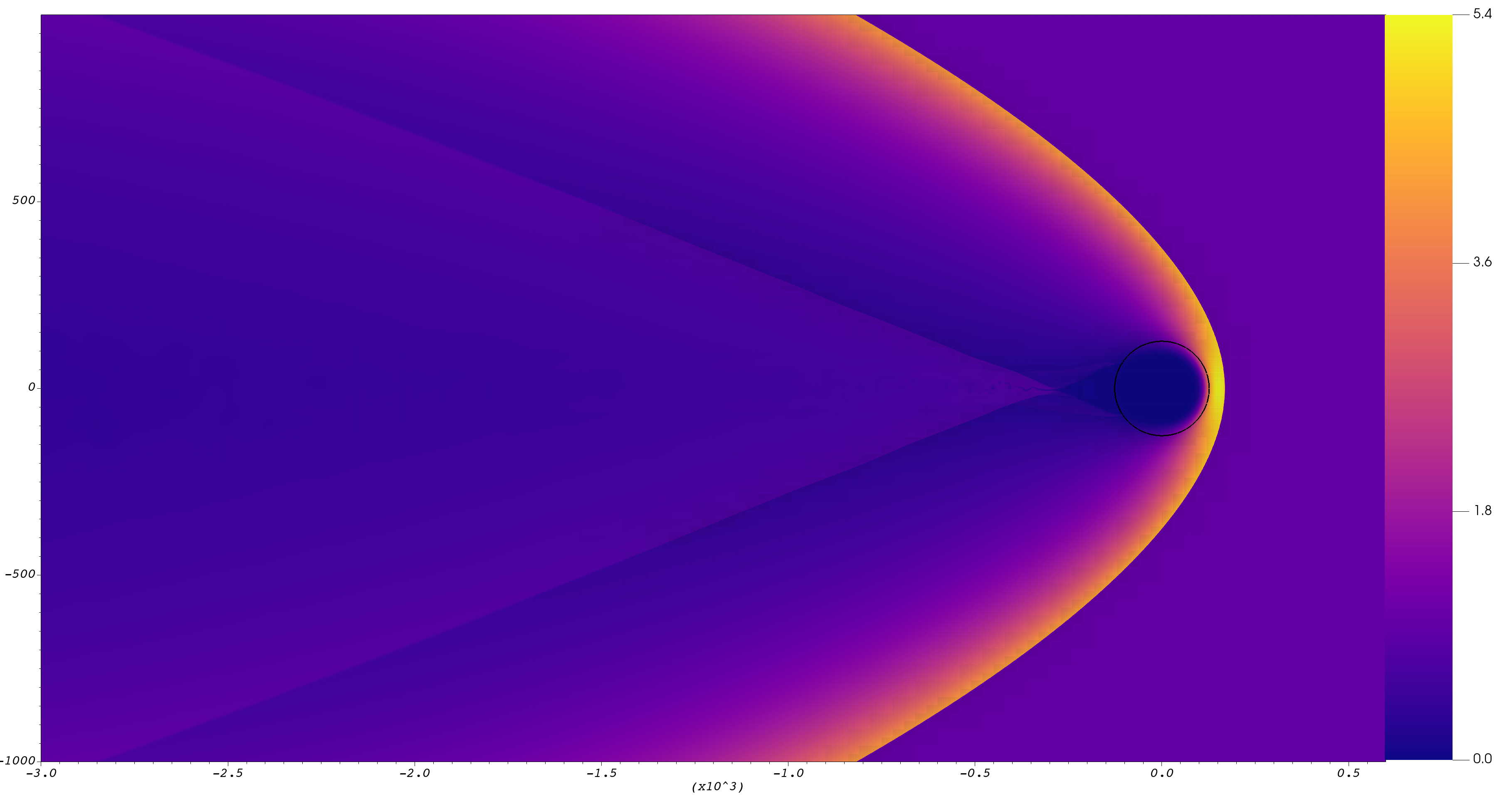}
    \caption{Snapshot of a stabilized simulation showcasing the primitive rest-mass 
             density $\rho$. The upstream temperature is $\Theta_1 = 10^{-6}$, the
             warp bubble speed is $v = 0.75c$ and the warp bubble radius is $R = 100$. The approximate location of the warp bubble wall is shown by the black circle near the shock nose.}
    \label{fig:enden}

    \vspace{.5em}

    \includegraphics[width=\textwidth]{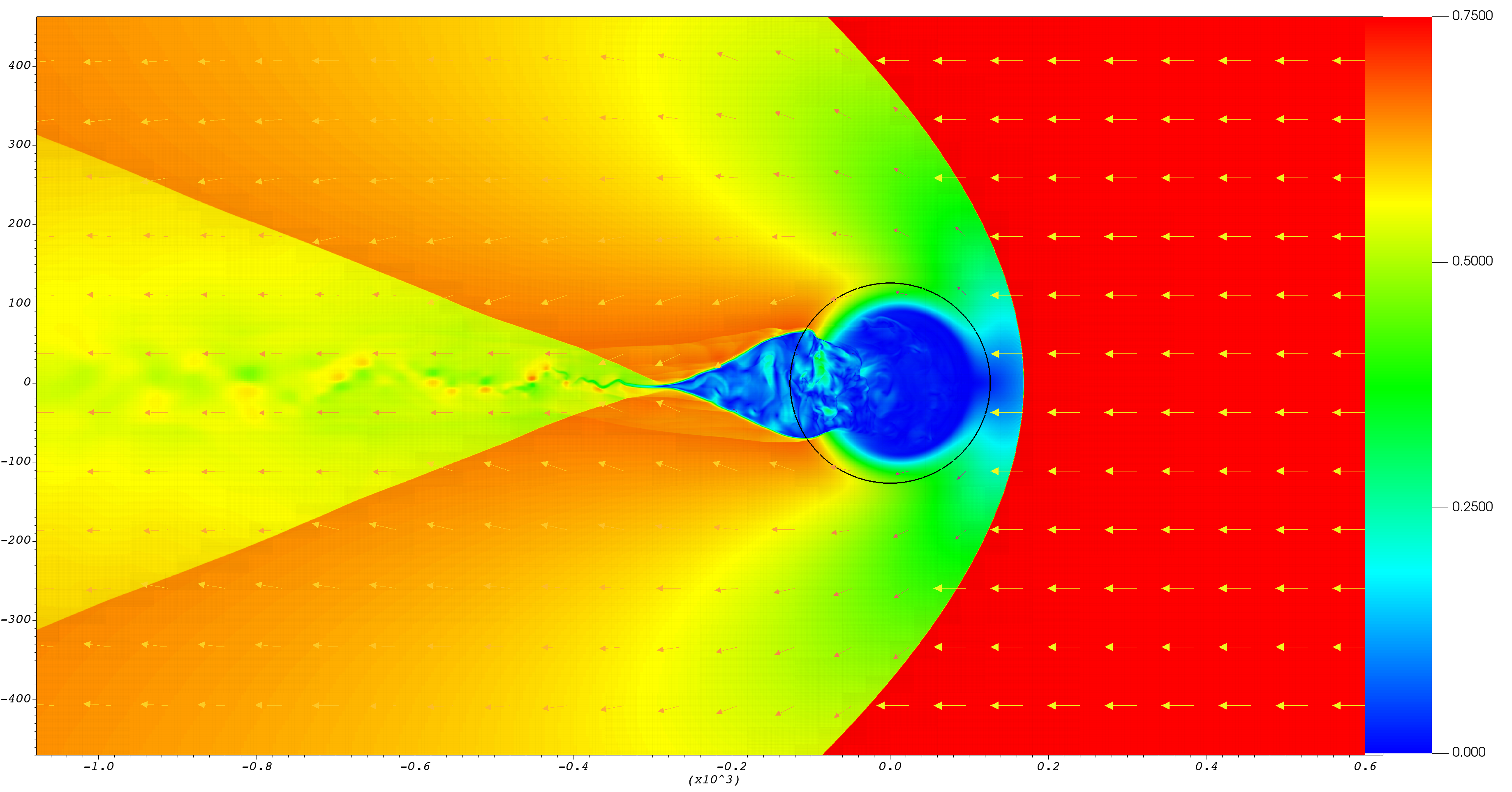}
    \caption{Snapshot of a stabilized simulation showcasing the fluid velocity norm in the
             warp bubble comoving frame. The upstream temperature is $\Theta_1 = 10^{-6}$,
             the warp bubble speed is $v = 0.75c$ and the warp bubble radius is $R = 100$. The approximate location of the warp bubble wall is shown by the black circle near the shock nose. The stagnation zone is clearly displayed between the bubble wall and the tip of the shock zone, denoted by the blue zone in front of the black circle. Overlaid are the 2D vectors of the velocity.}
    \label{fig:vnorm}
\end{figure*}

The final snapshot from one of the simulations is shown in Fig.~\ref{fig:enden} and Fig.~\ref{fig:vnorm}. The energy density plotted in Fig.~\ref{fig:enden} is the primitive energy density, recovered from the evolved variables using the scheme in Sec.~\ref{subsec:primitives}. The simulation ran for $t = 10000$ (code units), well past the convergence to a steady-state formation. The norm of the fluid velocity in the frame of the warp bubble is shown in Fig.~\ref{fig:vnorm}, computed using the equations in Sec.~\ref{subsec:dynamicalequatios}. There are several notable features to point out in this figure. The first is the very clearly defined stagnation zone between the warp bubble wall (represented by the black circle) and the nose of the shock zone, discussed in Sec.~\ref{subsec:stagnationandjumpconditions}. The second is the fluid being carried along in the interior of the warp bubble, as expected.  In the comoving frame, the metric shift field vanishes in this region, hence the occupying fluid is pulled along with the warp bubble. The third is the outflowing fluid away from the stagnation zone and around the warp bubble.

\begin{figure*}
    \centering
    \includegraphics[width=0.8\textwidth]{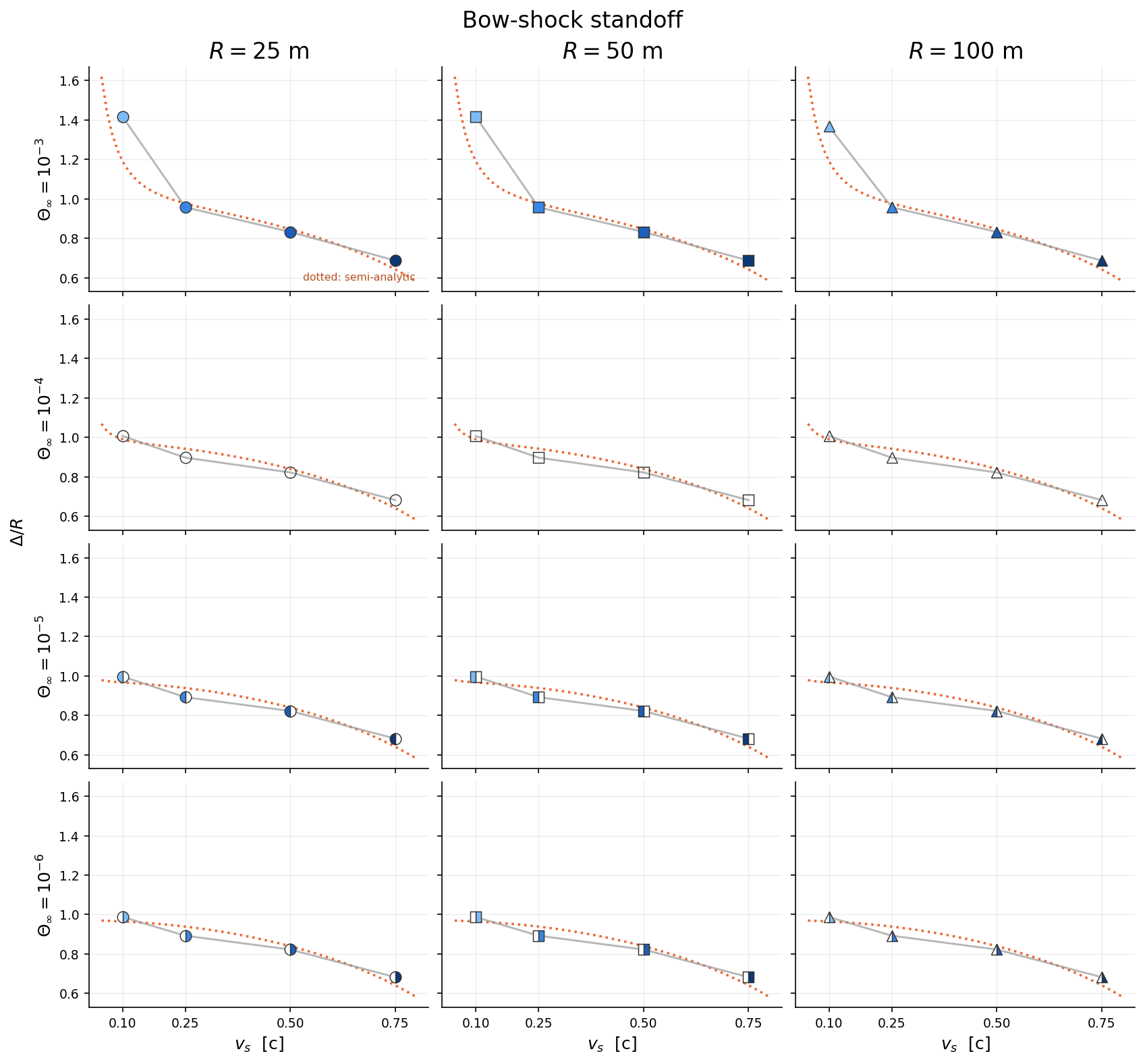}
    \caption{Fractional shock-front standoff relative to the bubble radius, Eq.~\eqref{eq:shockstandoff}. Simulated data is displayed by the individual circles or squares, while the analytical estimate, derived from the RH equations, is shown by the dotted line. Location in the grid of plots denotes the upstream temperature $\Theta_1$ (rows) and the bubble radius $R$ (columns). Each individual plot is the fractional standoff versus the bubble velocity.}
    \label{fig:standoff}
\end{figure*}

Fig.~\ref{fig:standoff} shows the shock standoff distance relative to the bubble radius. This figure, as with the grid plots that follow, is organized by upstream temperature $\Theta_{1}$ and bubble radius $R$, in the (row, column)-directions, respectively. Each plot shows the standoff distance versus the bubble velocity $v_s$. The discrete circle and square points are the post-processed simulation data and the dotted lines show the analytical results derived in Sec.~\ref{subsec:stagnationandjumpconditions}. The $O(1)$ constant is calibrated against all runs at once such that a single constant feeds into all analytic predictions for the standoff distance. As expected, the standoff distance depends heavily on the warp bubble velocity. Larger velocities correspond to a closer, more condensed shock front. On the other hand, the shock front is anti-correlated with the upstream temperature, which is also expected. Smaller upstream temperature corresponds to a smaller pressure term relative to the density, allowing the fluid to become more condensed and build-up closer to the bubble wall.

\begin{figure*}
    \centering
    \includegraphics[width=0.8\textwidth]{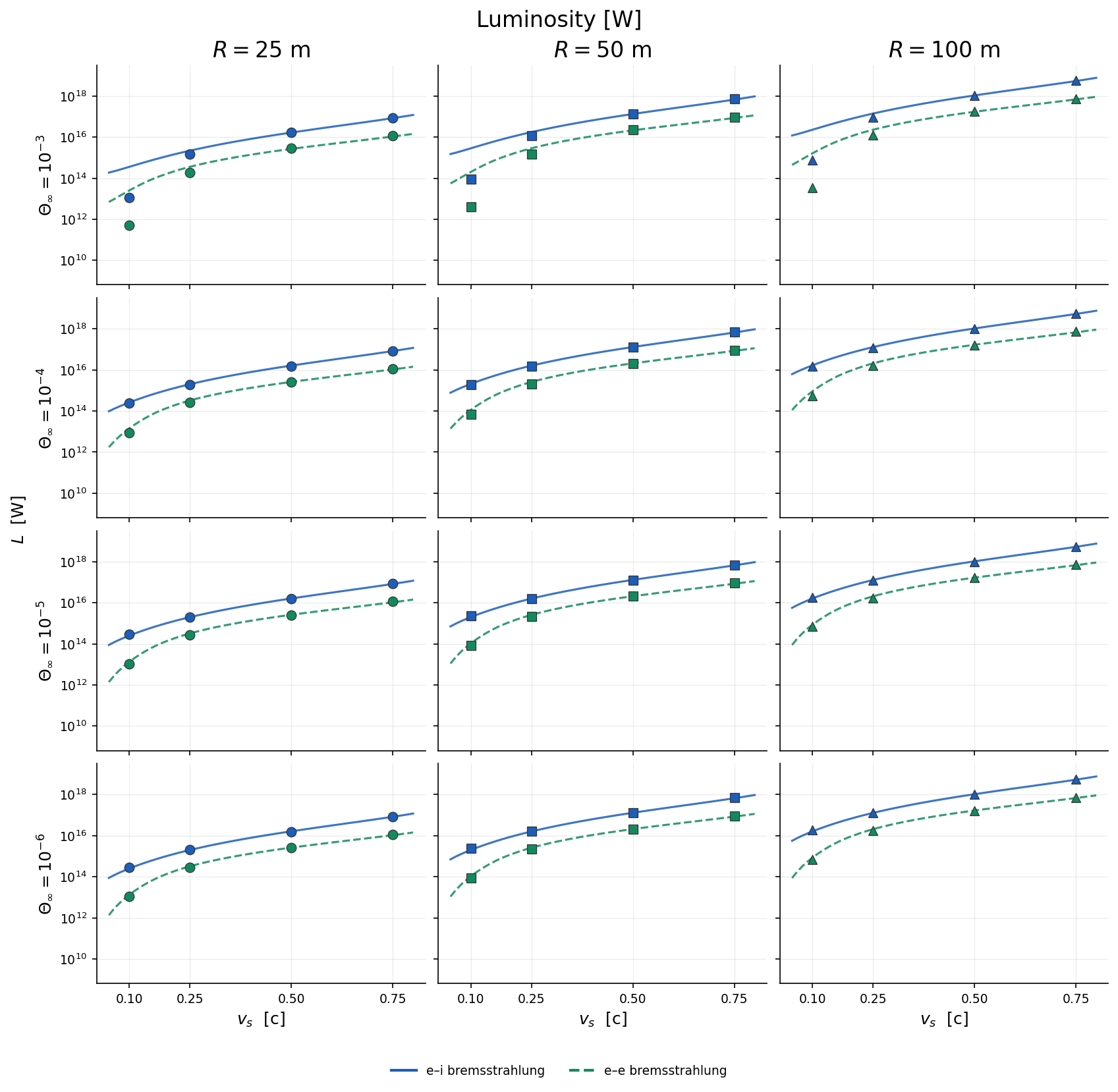}
    \caption{Total luminosity from $e^-e^-$ and $e^-i$ bremsstrahlung emissions. Simulated data is displayed by the individual circles or squares, while the analytical estimate is shown by the dotted line. Location in the grid of plots denotes the upstream temperature $\Theta_1$ (rows) and the bubble radius $R$ (columns).}
    \label{fig:Lchannels}
\end{figure*}

\begin{figure*}
    \centering
    \includegraphics[width=0.68\textwidth]{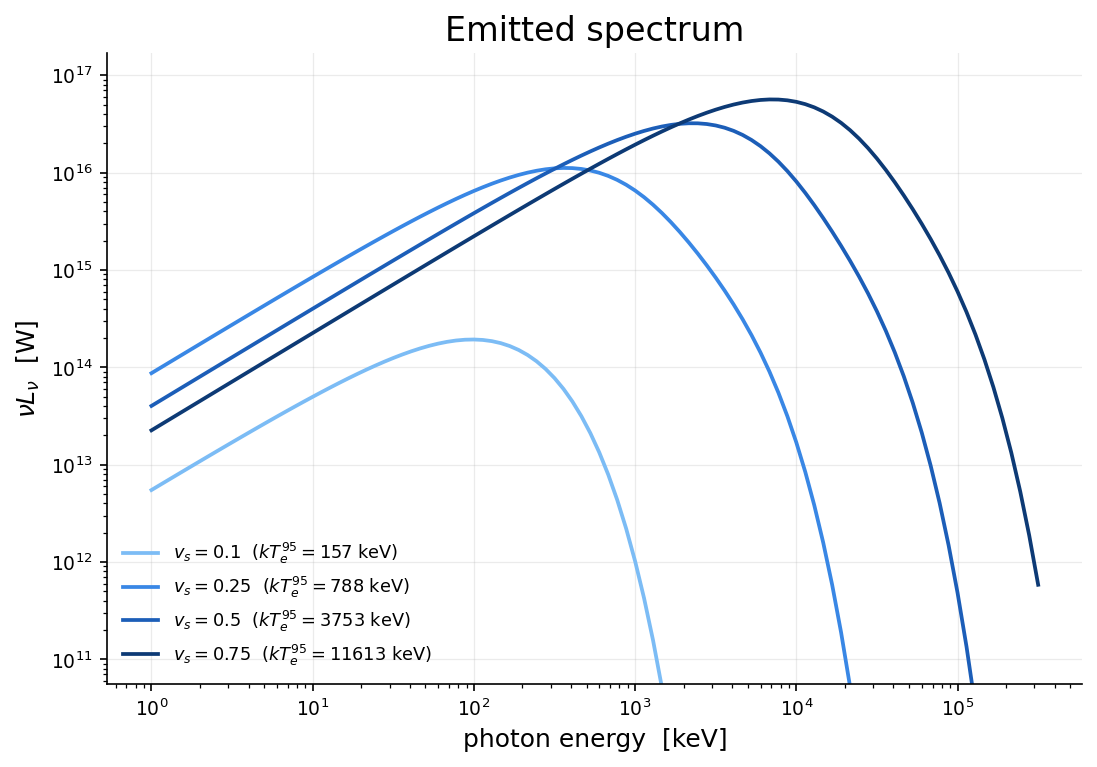}
    \caption{Displayed is the $e^-i$ bremsstrahlung spectrum and the given velocities and hence temperatures. The vast amount of the emitted photons are gamma to hard-gamma photons. Hard gamma emissions dominate the higher warp bubble velocity scenarios.}
    \label{fig:spectrum}
\end{figure*}

Fig.~\ref{fig:Lchannels} displays the total luminosity across all frequencies emitted from the shock zone from both $e^-e^-$ bremsstrahlung and $e^-i$ bremsstrahlung emissions. The total shock volume can extend over a great range, much larger than could be consistently simulated using the standard numerical methods. Therefore, the quoted luminosities are from the shock zones within $15R$ of the warp bubble center. These quoted luminosities are therefore a lower bound on the \emph{total} luminosity of the entire interaction, hence representing the luminosity of the near-nose shock zone. As expected, the $e^-i$ emissions channel completely dominates the total power. The total luminosity across all runs ranges from $10^{13}$ to $10^{19}$ watts, or 10's of terawatts to tens of exawatts. An observer at the surface, assumed to be $100$ km away, can estimate the brightness observed from such an object by taking into account atmospheric absorption and re-emission. The emitted spectrum of Eq.~\eqref{eq:emissionpower} from the shock zone is displayed in Fig.~\ref{fig:spectrum}. The dominant emission is gamma to hard-gamma photons. The spectrum observed by a surface observer can be estimated using the standard atmospheric absorption line. Assuming the interaction takes place at $100$ km and all the emission is from a point source, then the unattenuated flux reaching the surface is
\begin{equation}
    F_{\text{geom}} = \frac{L}{4 \pi d^2}
\end{equation}
which ranges from $\sim 80 \frac{\mathrm{W}}{\mathrm{m}^2}$ at the lower velocities, to $\sim 10^8\,  \frac{\mathrm{W}}{\mathrm{m}^2}$ at $v = 0.75$. The source spectrum itself is free-free emission, $L = L_0 e^{-\frac{h \nu}{k_B T_e}}$. The frequency-dependent atmospheric transmission behaves as $T(\nu) = e^{-\tau_{\text{atm}}(\nu)}$, where $\tau_{\text{atm}} = \frac{\mu}{\rho}(\nu) \Sigma$ is the atmospheric column optical depth, $\frac{\mu}{\rho}(\nu)$ is the frequency-dependent mass attenuation coefficient, and $\Sigma$ is the atmospheric mass column. Standard databases sort the frequency-dependent optical depth into distinct regions. The radio window ($0.1 - 197\, \mu \mathrm{eV}$) has very low optical depth, so $T \simeq 1$. The optical band ($1.8 - 3.9\,  \mathrm{eV}$ ) interacts with aerosols and the ozone, yielding $T_{\text{opt}} \sim 0.5 - 0.7$ with a naive optical depth around $\tau \sim 0.1$. The UV to soft gamma region ($3.9\,   \mathrm{eV} - 50\, \mathrm{MeV}$) is massively attenuated by photo-absorption, yielding $T \sim 0$. Deep gamma ($>50\,  \mathrm{MeV}$) is also highly suppressed, yielding $T \sim 0$~\cite{Bodhaine1999, xcom}. 

Since the emitted spectrum is mostly dominated by the $100$s of $\mathrm{keV}$ to $1-100\, \mathrm{MeV}$  photons, most of the energy goes into direct absorption and heating of the atmosphere. Since the luminosity is approximately isotropic, only about half is actually deposited into the upper atmospheric layers. The rest is radiated away into space. Note that this would change when the warp bubble enters the deep atmosphere near the surface, in which case the present study places a \emph{lower} bound on the total observed luminosity. The deposited flux across the lower hemisphere, centered on the warp bubble, is then
\begin{equation}
    F_{\text{dep}} = \frac{L/2}{4 \pi d^2}
\end{equation}
where $d$ is the distance to the atmospheric layer that absorbs the high-energy photons, here taken to be the upper stratosphere, hence $d \approx 50$ km.
The deposited energy ionizes and heats the stratospheric layers, which re-radiate in channels the atmosphere is transparent to. The physics of cosmic radiation incident on the atmosphere fixes the efficiency of converting high-energy photons into UV and optical bands, which is determined by the measured air-fluorescence yield~\cite{Ave2008}. The efficiency of the fluorescence is approximately $\eta_{\text{fluo}} \sim 2\times 10^{-4}$ at $\sim 3.5\,  \mathrm{eV}/\text{photon}$ and $20-50$ kilometer altitude. Thus, the flux from the re-radiated glow is 
\begin{equation}
    F_{\text{glow}} = \eta_{\text{fluo}}F_{\text{dep}} T_{\text{strat}}
\end{equation}
evaluated at the stratospheric layers, since this is where the atmosphere becomes thick enough to absorb the high-energy photons. For the transmission coefficient of $T_{\text{strat}} \approx 0.6$, the flux through the layer then ranges from $0.007 \frac{\mathrm{W}}{\mathrm{m}^2}$ for $v_s = 0.1c$ and $R = 25$m up to $7460 \frac{\mathrm{W}}{\mathrm{m}^2}$ at $v_s = 0.75c$ and $R = 100$m. Thus, the dominant visual effect will be a highly illuminated stratospheric layer below the shock cone with a small fraction of the total luminosity in the form of direct optical photons. These numbers are all order-of-magnitude. For a more precise calculation, one would need to directly compute the microphysics of the interaction and how the emitted spectrum propagates through the atmospheric layers, which is beyond the scope of this study.

\begin{figure*}
    \centering
    \includegraphics[width=0.68\textwidth]{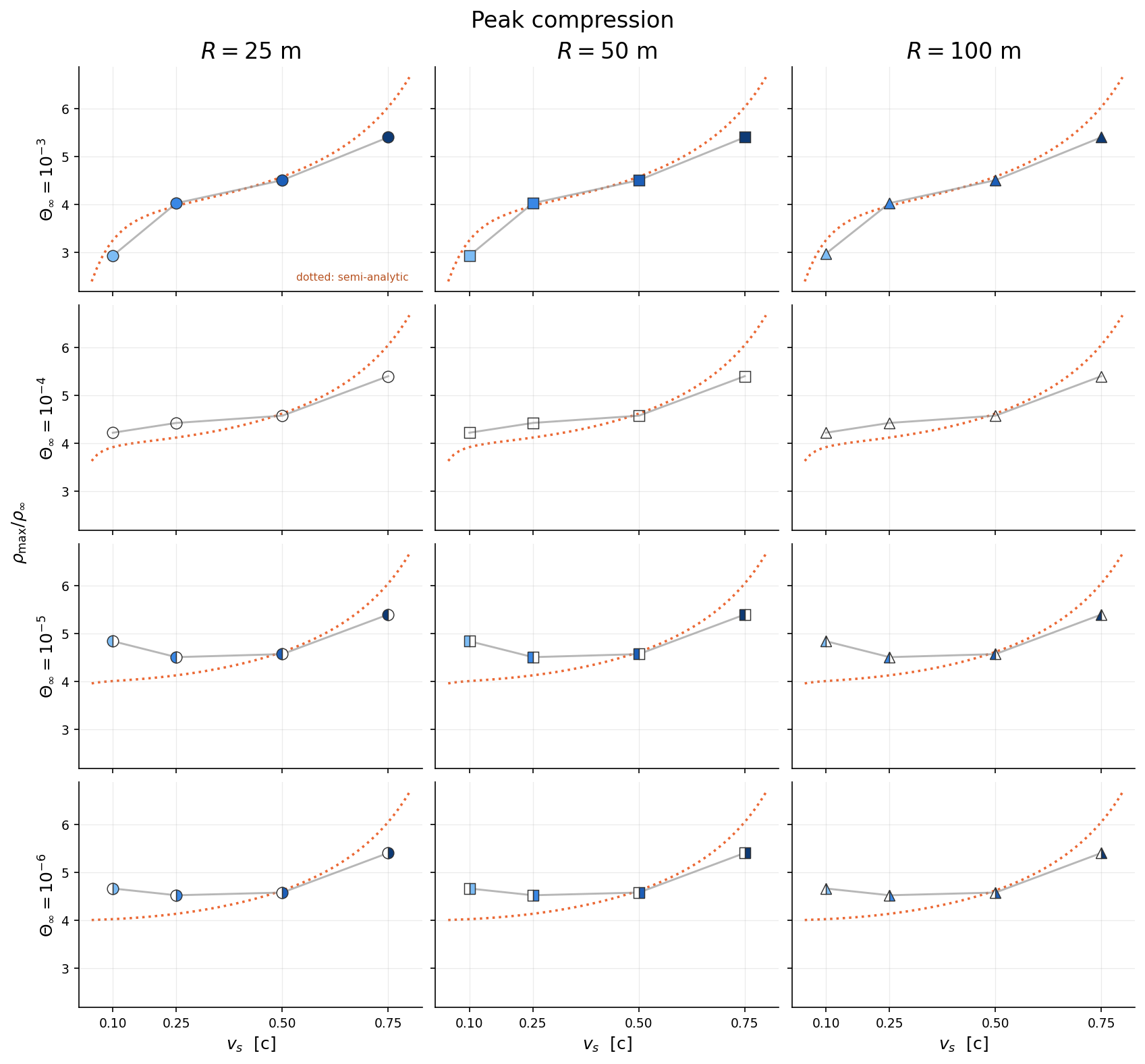}
    \caption{Maximum compression ratio of the post-shock fluid to the upstream fluid, $\frac{\rho_2}{\rho_1}$. Simulated data is displayed by the individual circles or squares, while the analytical estimate is shown by the dotted line. Location in the grid of plots denotes the upstream temperature $\Theta_1$ (rows) and the bubble radius $R$ (columns). The dotted curves are the analytical predictions directly from the RH conditions.}
    \label{fig:compression}
\end{figure*}

The compression ratio $\frac{\rho_2}{\rho_1}$ from Eq.~\eqref{eq:compratio} is displayed in Fig.~\ref{fig:compression}. The ratio is computed along the direction of travel, through the stagnation point, and measures the maximum density relative to the incoming fluid, $\frac{\rho_2}{\rho_1}$, along this line. 

To confirm the finite-difference scheme converges to a physical solution, two error-estimates are computed. The first is the steady rest-mass residual. Rest-mass conservation is $\nabla_{\mu}(\rho u^{\mu}) = 0$. In the given metric, this results in the continuum constraint
\begin{equation}
    \partial_x (\rho u^x) + \partial_y(\rho u^y) = 0\,.
\end{equation}
To make it dimensionless and mesh-comparable, it is normalized by the local mass-flux magnitude carried across one cell. Thus, the figure of merit is
\begin{equation}
    \epsilon_D := \frac{|\partial_i (\rho u^i) \Delta x|}{|\rho \vec{u}|}\,.
\end{equation}
At a captured shock, the flux turns over inside the $O(1)$-cell numerical layer, so $\epsilon_D \rightarrow O(1)$ by nature of the finite difference. The importance lies therefore in the smallness of $\epsilon_D$ in the post-shock zone.

The second convergence factor is the relativistic Bernoulli invariant. Using conservation of the stress-energy tensor, together with the temporal killing vector and assuming a steady-flow state, one finds that
\begin{equation}
    u^{\mu} \partial_{\mu}(h u_t) = 0,
\end{equation}
implying that $h u_t$ is constant along the streamlines. Let $B \equiv -h u_t$. With a uniform upstream state, every streamline enters the domain carrying the same $B$, so let 
\begin{equation}
B_{\text{ref}} = \underset{\text{\tiny upstream}}{\text{median}}\left( B\right).
\end{equation}
The figure of merit is then the deviation from this
\begin{equation}
    \delta B = \frac{ B - B_{\text{ref}}}{B_{\text{ref}}} = 0 \;,
\end{equation}
which is purely algebraic so does not require finite differences across a mesh.

\begin{figure*}
    \centering
    \includegraphics[width=\textwidth]{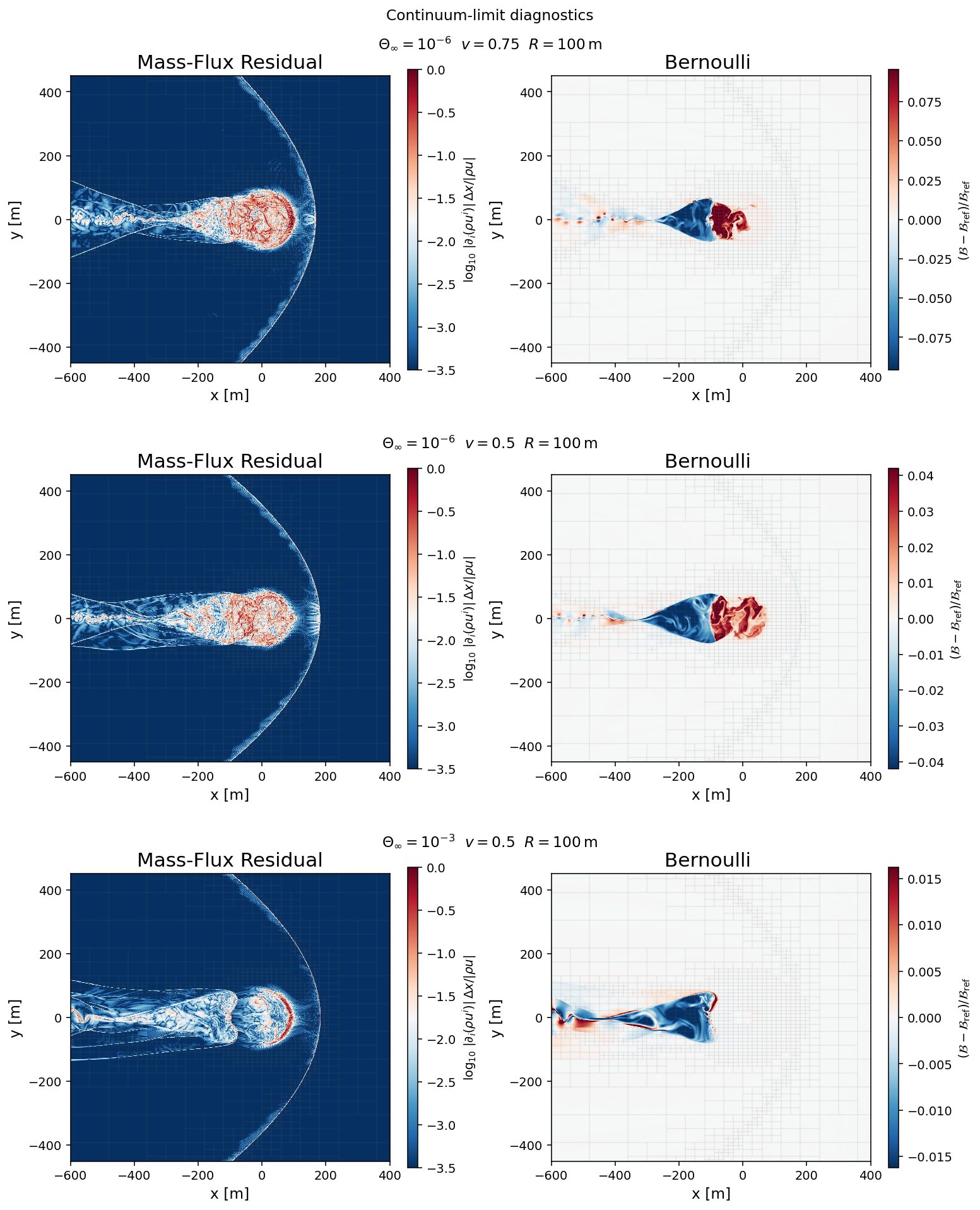}
    \caption{Convergence figures of merit for 3 chosen simulation runs, with varying levels of computational challenge. The most important regions in which the figures of merit hold are the shock zones, where the most important physics is occurring. The figures of merit are \emph{not} expected to necessarily hold in the highly rarefied wake or the bubble interior, which have very small density values and thus magnified error.}
    \label{fig:convergence}
\end{figure*}

Fig.~\ref{fig:convergence} shows the convergence analysis on the final snapshot of a select number of simulations. The chosen simulations span the coldest to the hottest upstream temperatures and the slowest and fastest warp bubble velocities. As expected, the shock front converges very nicely with vanishing Bernoulli error and well-converged mass-flux residuals. The most difficult part of the grid to numerically simulate is the highly rarefied zones, consisting of the warp bubble interior and the rarefied wake. In these regions, the density is extremely small, c.f. Fig.~\ref{fig:enden}, which thus amplifies the error. These regions are not interesting since their dynamics do not propagate to the shock zone, which is where the interesting quantities come from.

\bigskip
\section{Discussion and Interpretations} \label{sec:discussion}

There are several interesting conclusions one could draw from the results of this study. Primarily, the observational signatures of a high-altitude relativistic warp drive present a novel signal. An observer on the surface will witness two distinct signals. First, a direct high-energy signal comes from the shock zone itself. The bremsstrahlung spectrum that survives the atmospheric penetration will manifest as an extremely bright source of light in the sky, ranging from fractions of the solar radiance to hundreds of times it. Secondly, the extremely energetic photons penetrate the upper layers and only become absorbed by the lower atmospheric layers. The extreme energy gets deposited in these layers and re-radiates in the observable bands. This will manifest as a bright atmospheric glow directly below the shock zone. Due to the extremely energetic interaction, the atmospheric glow can also completely dominate the surface solar flux by an order of magnitude.

\section{Conclusion} \label{sec:conclusion}

We simulated the consequences of an Alcubierre-type warp drive moving relativistically through the Earth's atmosphere. Our results show that the relativistic motion of aircraft-scale spacetime bubbles through air generates luminosities in excess of a terawatt. The inferred luminosities exceed by many orders of magnitude the values detected for UAP~\cite{Watters,Knuth}. Nonetheless, a sufficiently advanced civilization that visits Earth using warp drive techniques will leave a unique observational signature, should they enter the atmosphere at such high velocities. These results further constrain warp drive interpretations of UAP within the terrestrial environment.

The results of this study are strictly for zero ADM mass warp drives, such as the Alcubierre metric~\cite{alcubierre1994warp}, the Van den Broeck metric~\cite{vandenbroeck1999warp}, and the Lentz metric~\cite{lentz2021breaking}. It has been shown in~\citet{bobrick2021introducing} that generic warp drive spacetimes have non-zero ADM mass and thus a fundamentally distinct asymptotic structure than the zero ADM mass variants. Non-zero ADM mass spacetimes would behave differently from those in the current study, due to the different behavior of the curvature. For example, the standoff distance~\eqref{eq:shockstandoff} assumes a flat (or exponentially flat) spacetime outside a given characteristic distance $R$. A non-zero ADM mass spacetime would instead have non-zero curvature throughout the entire simulation space and thus a modified shock standoff distance. Future studies will explore this region of the parameter space, including sub-relativistic velocities.

In the non-relativistic velocity regime and above the relativistic-enhancement knee, the calculated luminosity scales with the cube of the radius $R$ of the spacetime bubble and the square of the ambient density, $L \propto \rho_1^2 R^3$. The $R^3$ is due to the optical properties of the shock zone, where the entire emitting volume contributes to the observed emissions and not just a single radiating surface, combined with the standoff distance growing linearly with $R$, Eq.~\eqref{eq:shockstandoff}. Additionally, the velocity dependence is not a single power. Above the relativistic-enhancement knee at $k_B T_e \simeq 0.2\, \mathrm{MeV}$, where the enhancement factor of Eq.~\eqref{eq:emissionpower} turns $\sqrt{T_e}$ into $T_e^{3/2}$, the luminosity scales with the cube of the velocity, $L \propto v_s^3$, while below the knee it reverts to $L \propto v_s$. 

Using this scaling, one can estimate how faint such an object could be made. Lowering the luminosity to the scale of $1$ watt would require around $13$ orders of magnitude of dimming below the faintest parameter configuration simulated here. Slowing the bubble drops it below the enhancement knee, where the cube law flattens to $L \propto v_s$, while shrinking it terminates at the continuum floor, $R \gtrsim 14\, \mathrm{m}$. Minimizing the luminosity subject to both constraints places the floor near $10^3$ watts for a micron-scale bubble at sea-level density moving at several times the speed of sound. This floor, however, bounds the present calculation rather than the object itself. Both limits mark the edge of the emission mechanism: below the continuum floor the flow becomes free-molecular and no compressed shell forms, while below the ionization threshold the shocked air retains its electrons and the free-free channel of Eq.~\eqref{eq:emissionpower} turns off. A sufficiently compact or slow warp drive would therefore not shock the atmosphere, and would not glow by this mechanism at all. The constraints derived here bound relativistic, aircraft-scale transits; they leave the warp drive interpretation of UAP open in the compact and subsonic regimes, whose signatures are set by kinetic rather than hydrodynamic physics and remain to be computed.

\section{Acknowledgements} 
The computations presented in this work were performed on the FASRC Cannon cluster, supported by the FAS Division of Science Research Computing Group at Harvard University. This research was supported in part by the Galileo Project at Harvard University. The authors also acknowledge Alexey Bobrick and Gianni Martire for their insightful comments and valuable feedback during the preparation of this manuscript.

\clearpage
\bibliographystyle{apsrev4-2}
\bibliography{bibliography}

\end{document}